# MULTICHANNEL ALGORITHM BASED ON GENERALIZED POSITIONAL NUMERATION SYSTEM

*Copyright 2006 © **Alexandre Lavrenov***

*Abstract: This report is devoted to introduction in multichannel algorithm based on generalized numeration notations (GPN). The internal, external and mixed account are entered. The concept of the GPN and it's classification as decomposition of an integer on composed of integers is discussed. Realization of multichannel algorithm on the basis of GPN is introduced. In particular, some properties of Fibonacci multichannel algorithm are discussed.*

A basis of this article is simple arithmetic equality - decomposition of an integer on composed of integers. Using this decomposition the interrelation of the various parties of the science is shown. The plan of this article is following. In the beginning communication of decomposition of an integer on composed of integers with classification of the generalized positional notations (GPN) is considered. Then concepts of the internal, external and mixed account are entered. Bit transformation of the initial alphabet to set of other alphabets is discussed, using the decomposition of an integer specified above. Realization of multichannel algorithm on the basis of GPN is considered. In particular, some properties Fibonacci multichannel algorithm are discussed.

At the beginning let's give the formulation of a problem - decomposition of an integer on composed of integers. Procedure easy enough, but demanding detailed elaboration. One of the first specifications concerns quantities composed. Really, at decomposition of integer $N$ on $n$ composed there is finite set of decisions. In Table 1 at small values of number possible variants are shown all. From the commutativity of addition in Tables 2,3 are given independent variants of decisions for next values of number $N$. Also last line of each table value of product of all composed is given, and its maximal size is allocated by more dark background of a corresponding cell.

| Integer $N$ \ composed | 1 | 2 | | 3 | | | 4 | | | | | 5 | | | | | | | | | | | | | | | |
|---|---|---|---|---|---|---|---|---|---|---|---|---|---|---|---|---|---|---|---|---|---|---|---|---|---|---|---|
| 1 | 1 | 1 | | 1 | 2 | | 1 | 1 | 2 | 3 | | 1 | 1 | 2 | 1 | 1 | 2 | 3 | 4 | 1 | 1 | 1 | 2 | 2 | 3 | 1 | 1 | 1 | 2 | 1 |
| 2 | | | 1 | 2 | 1 | | 1 | 3 | 2 | 1 | 1 | 2 | 1 | 1 | 4 | 3 | 2 | 1 | | 1 | 2 | 3 | 1 | 2 | 1 | 1 | 1 | 2 | 1 | 1 |
| 3 | | | | | | 1 | | | 2 | 1 | 1 | 1 | | | | | 3 | 2 | 1 | 2 | 1 | 1 | 1 | 2 | 1 | 1 | 1 |
| 4 | | | | | | | | | | | | 1 | | | | | | | | | | | | 2 | 1 | 1 | 1 | 1 |
| 5 | | | | | | | | | | | | | | | | | | | | | | | | | | | | | | 1 |
| Product | 1 | 1 | 2 | 2 | 1 | 3 | 4 | 3 | 2 | 2 | 1 | 4 | 6 | 6 | 4 | 3 | 4 | 3 | 4 | 4 | 3 | 2 | 2 | 2 | 2 | 1 |

Table 1. Every possible variants of decomposition of integer $N$ on $n$ composed.

From the resulted tables it is visible, that such decomposition are similar to record of number in the certain notation with the basis not above value $(N-n)$. Such sensation is not casual. Therefore the process of the account should be considered in more details and attentively.

The account has arisen from practical activities of the person on ordering and the account of subjects. Therefore let's choose $N$ objects, which nature is not important at present, and we shall order them. In other words, let's them arrange physically, for example, from itself forward one behind another. Establishing isomorphism or unequivocal conformity between the given objects and numbers or a numerical axis, there is a following picture. The choice of one any object will be a choice of any number. Hence, touching all our objects, we actually recalculate them. We shall name such account *internal* because of a choice as unit of the account one object.

| *Integer N \ composed* | 6 | | | | | | | | | | 7 | | | | | | | | | | | | | |
|---|---|---|---|---|---|---|---|---|---|---|---|---|---|---|---|---|---|---|---|---|---|---|---|---|
| *1* | 1 | 2 | 3 | 1 | 1 | 2 | 1 | 1 | 1 | 1 | 1 | 2 | 3 | 1 | 1 | 1 | 2 | 1 | 1 | 1 | 1 | 1 | 1 | 1 |
| *2* | 5 | 4 | 3 | 1 | 2 | 2 | 1 | 1 | 1 | 1 | 6 | 5 | 4 | 1 | 2 | 3 | 2 | 1 | 1 | 2 | 1 | 1 | 1 | 1 |
| *3* | | | | 4 | 3 | 2 | 1 | 2 | 1 | 1 | | | | 5 | 4 | 3 | 3 | 1 | 2 | 2 | 1 | 1 | 1 | 1 |
| *4* | | | | | | | 3 | 2 | 1 | 1 | | | | | | | | 4 | 3 | 2 | 1 | 2 | 1 | 1 |
| *5* | | | | | | | | | 2 | 1 | | | | | | | | | | | 3 | 2 | 1 | 1 |
| *6* | | | | | | | | | | 1 | | | | | | | | | | | | | 2 | 1 |
| *7* | | | | | | | | | | | | | | | | | | | | | | | | 1 |
| Product | 5 | 8 | 9 | 4 | 6 | 8 | 3 | 4 | 2 | 1 | 6 | 10 | 12 | 5 | 8 | 9 | 12 | 4 | 6 | 8 | 3 | 4 | 2 | 1 |

Table 2. Independent variants of decomposition *N=6* and *N=7* on *n* composed.

| *Integer N \ composed* | 8 | | | | | | | | | | | | | | | | | | | | |
|---|---|---|---|---|---|---|---|---|---|---|---|---|---|---|---|---|---|---|---|---|---|
| *1* | 1 | 2 | 3 | 4 | 1 | 1 | 1 | 2 | 2 | 1 | 1 | 1 | 1 | 2 | 1 | 1 | 1 | 1 | 1 | 1 | 1 |
| *2* | 7 | 6 | 5 | 4 | 1 | 2 | 3 | 2 | 3 | 1 | 1 | 1 | 2 | 2 | 1 | 1 | 1 | 1 | 1 | 1 | 1 |
| *3* | | | | | 6 | 5 | 4 | 4 | 3 | 1 | 2 | 3 | 2 | 2 | 1 | 1 | 2 | 1 | 1 | 1 | 1 |
| *4* | | | | | | | | | | 5 | 4 | 3 | 3 | 2 | 1 | 2 | 2 | 1 | 1 | 1 | 1 |
| *5* | | | | | | | | | | | | | | | 4 | 3 | 2 | 1 | 2 | 1 | 1 |
| *6* | | | | | | | | | | | | | | | | | | 3 | 2 | 1 | 1 |
| *7* | | | | | | | | | | | | | | | | | | | | 2 | 1 |
| Product | 7 | 12 | 15 | 16 | 6 | 10 | 12 | 16 | 18 | 5 | 8 | 9 | 12 | 16 | 4 | 6 | 8 | 3 | 4 | 2 | 1 |

Table 3. Independent variants of decomposition *N=8* on *n* composed.

However it is possible to act in another way. Having chosen for a unit of measure all set from ours *N* objects, it is possible to receive or empty set (*(N+1)* a state) or all set (*(N+2)* a state), i. e. there are only two values or states. Let's name such account *external* as it is the account not inside of initial set, and actually the account of the set and external in relation to objects. Usually in practice the given two accounts mix. Record of number is carried out by means of symbols of any alphabet where almost always include zero as (*(N+1)*) a state. Thus, as a result of the described procedure of the account by means of *N* objects it is possible to describe all (*N+2*) states.

Let's complicate our analysis [1]. We shall consider a situation when we have taken some such ordered sets or, on the contrary (that is in common equivalent), have divided initial set *A* on *n* subsets $A_n$ generally with unequal quantity of objects. i. e. in language of sets it we shall express so

$$A = \sum_n A_n . \qquad (1)$$

In each its elements $A_n$ are ordered by means of the alphabet $a_n$, and their total or capacity (power) of the alphabet $a_n$ is $P(a_n) = p_n$. Hence, any set of single element of each set can be written down as follows:

$$\{X\} = \sum_{m=1}^{m=n} a_m A_m \equiv a_n a_{n-1} \ldots a_m \ldots a_2 a_1. \quad (2)$$

For reception of new qualitative result, we shall do following procedure. We shall order both objects inside of subsets, and the chosen subsets. Objects can be presented physically located not only further or more close (ordering of objects inside of a subset), but also placed on the right or to the left of itself as representatives of different subsets (ordering of subsets). In other words, we have bidimentional space of ordering of objects. For the description of a maximum quantity of states we shall enter a following rule-conformity. Let (N+2) a state of the previous subset (as the subset) will be unit of a following subset. It allows at transition from one subset to another to change unit of the account for number of objects of the previous subset plus one (due to (N+1) or a zero state). Thus, choosing on one representative from each subset, we have an opportunity to describe the quantity of states much more exceeding total of objects. Representation of this number in the form of decomposition on *n* composed, each of which can be various, actually corresponds to the division of initial set described above from *N* objects on *n* the ordered subsets. In other words, we have the certain system of the account or notation. We shall name its generalized notation (GPN). All discussed in literature GPN are described in our approach. If to choose accordingly $A_m$ as $b^m$, $m!$ or $C_{n-1}^m$, we shall receive $b$ - radix polynomial, factorial and binomial GPN. As example GPN we shall offer in a role of radix the generalized $k$ - deformed Fibonacci matrixes $Q_k = \begin{bmatrix} \vec{c} & c_k \\ I_{k-1} & 0 \end{bmatrix}$, where $\vec{c} = diag(c_1, c_2, \ldots c_{k-1})$ and $c_k$ - parameters of deformation.

Decomposition on certain composed the general number of objects allows to lead somewhat classification GPN. From above-stated construction GPN it is clear, that capacity $N = P_A$ of set $A$ is equal to the sum of capacities $P_{A_n}$ of subsets $A_n$:

$$P_A \equiv \sum_{m=1}^{m=n} P_{A_m}. \quad (3)$$

In other words, there is the connection of GPN classification with decomposition $P_A$ on sum of composed $P_{A_n}$. In this direction there is an interesting problem about the most effective decomposition of such number. In another way it can be formulated as follows - at the fixed value of number *N* to find its such decomposition on composed that their product was maximal. In our case to it will correspond such GPN which will describe the maximal number of states at the fixed value of all objects. At splitting into two numbers composed by effective decomposition there will be its splitting half-and-half. Hence, the most effective GPN will be GPN with equal number of objects which will describe to us corresponding number of states. If to extrapolate the given result on the general case the choice of mankind as daily GPN as GPN with equal quantity of objects in subsets does not look casual. Detailed consideration of a situation in case of *n* composed for greater and any values of number *N* complex enough will not be resulted here again. As small acknowledgement told above it is possible to consider the result given in tables 1-3.

Let's return a little back and we shall recollect internal and external accounts for one set. By consideration of a situation with several subsets us gets out on their one representative, that actually means search or recalculation in an object way. In other words, the unit of measure or accounts is object and, accordingly, the account is internal. What corresponds in this case to the external account? Remember our rule: (N+2) a state of the previous subset (as a subset) is unit of a following subset. It means that we have the account external. Hence, in this case the account turns out mixed. Though unit of the account is object, but in the subsequent subset it represents all previous subset because of our rule-conformity. It can be not accepted. In this case each subset will have the unit of measure depending on a position in record of number. But the numerical axis one and, accordingly, exists interrelation between all units of measure.

Thus, we shall sum up on GPN:

*generalized positional notations represents the number* as $a_N a_{N-1} \ldots a_2 a_1$ and are determined by following items:

*GPN numerical function* defines a quantitative equivalent of considered number $a_N a_{N-1} \ldots a_2 a_1$ by following expression: $a_N a_{N-1} \ldots a_2 a_1 = \sum_{k=1}^{k=N} a_k R_k$;

*GPN alphabet* - $a_k$ (the binary alphabet will be used, i.e. $a_k \in B = \{0;1\}$);

*GPN weights* - $R_k$;

*Binary radix* – radix with $R_k = 2^{k-1}$;

*Binomial radix* - radix with binomial weights, i.e. $B_j = b_1 b_2 \ldots b_j \ldots b_r = \sum_{j=1}^{j=r} b_j C_{n-j}^{k-q_j}$;

*Fibonacci radix* – radix with $R_k = F(k)$, where $F(k)$ - Fibonacci numbers.

Let's consider now as it is possible to realize multichannel algorithm on the basis of GPN. For this goal let's recollect that the basis of MV2 algorithm [2] as the expressive example of multichannel crypto-algorithms began with the equation:

$$2^n = \sum_{k=1}^{k=n-1} 2^k + 2 = 2^1 + \ldots + 2^{n-1} + 2. \quad (4)$$

This equation shows that $N$-digit alphabet ($A_N$) can be divided into finite set of alphabets ($A_K$) with constant length of a tuple $1 \leq K \leq N-1$ plus two any elements from initial $N$-digit alphabet ($A_N$), i.e.

$$A_N = \sum_{K=1}^{K=N-1} A_K + a_N^1 + a_N^2, \quad (5)$$

where $a_K^i \in A_K$ and $i \in [1; 2^K]$.

For a reminder of the given algorithm we shall present a simple MV2 algorithm example based on $N$-digit tuple, where $N = 2$. There are two tables – Table 4 and Table 5. The first item represents isomorphism between a finite set of symbols and the two-digit alphabet ($A_{N=2} = A_2$).

| 1 | Symbols | Two-digit alphabet ($A_2$) |
|---|---------|----------------------------|
| 1 | a       | 00                         |
| 2 | b       | 01                         |
| 3 | c       | 10                         |
| 4 | d       | 11                         |

Table 4. Realization of two-digit alphabet ($A_2$).

The second item shows that two-digit alphabet ($A_2$) can be divided into finite set of alphabets ($A_K$) with constant length of a tuple $K = 1$ plus two any elements from initial two-digit alphabet ($A_2$), i.e.

$$A_2 = A_1 + a_2^1 + a_2^2, \quad (6)$$

where $a_2^i \in A_2$ and $i \in [1;2]$.

For the best understanding realization of MV2 algorithm is displayed for the some random file in Table 6 in according the Table [1] 5.

| [1] | Two-digit alphabet ($A_2$) | MV2 bit recording |
|---|---|---|
| 1 | 00 | 0 |
| 2 | 01 | 00 |
| 3 | 10 | 11 |
| 4 | 11 | 1 |

Table 5. Realization of MV2 bit recording for alphabets ($A_2$).

| Number of round for MV2 bit recording | Random file or core (symbol representation) | Random file or core (binary representation) | Flag |
|---|---|---|---|
| 0 (Initial) | a a d d b b | 00 00 11 11 01 01 | - |
| 1 | a d a a | 0 0 1 1 00 00 | 10 10 10 10 1 1 |
| 2 | b a | 0 1 0 0 | 10 10 10 10 |

Table 6. Realization of MV2 algorithm for some random file.

According to this MV2 bit recording of *N*-digit alphabet ($A_{N=2} = A_2$) the initial file is divided on two parts and any of them has no neither statistical and nor a semantic link with an initial file. The first part names a core (kernel) and appears from an initial file by bit recording (see Tables 5 and 6)). The second part (so-called flag) gives us the information on the length of a tuple *K* for each element. This information is the information on number of the used alphabet also. Because of obvious compression of an initial file in comparison with the core, all procedure can be repeated some times. It is a key moments of MV2 algorithm, which are shown above.

There is an opportunity to realize clones of MV2 algorithm [3-4] - for it is necessary to use instead of the equation (5) the following equations:

$$A_N = \sum_{K=1}^{K=N} \sum_{i=1}^{i=M_K} a_K^i ; \qquad (7a)$$

$$\sum_{K=1}^{K=N} M_K = 2^N , \qquad (7b)$$

where $M_K$ is quantity of the elements of alphabet ($A_K$).

Also equation (4) can be transformed to other, combinatorial form:

$$2^N = (1+1)^N = \sum_{k=0}^{k=N} C_N^K = C_N^0 + C_N^1 + \ldots + C_N^{N-1} + C_N^N . \qquad (8)$$

It shows that the *N*-digit alphabet ($A_n$) also can be divided into finite set of *N*-digit *K*-binomial alphabets ($B_{K,N}$) [5], i.e.

$$A_N = \sum_{K=1}^{K=N} B_{K,N} \equiv \sum_{k=1}^{k=N-1}(X_k + Y_k) + B_{N,N} ; \qquad (9)$$

$$B_{N,N} \equiv \{0;1\}.$$

This binomial multichannel algorithm was proposed in [6]. Continuing to generalize procedure of bit recording on other notations we shall consider equality (3) as equality (4). In other words, if to establish (install) isomorphism between elements $n$-digit GPN and elements $l$-digit GPN ($1 \leq l \leq n$) it is possible to enter the multichannel algorithm on the basis of GPN. In a kind of that generally $P_{A_i} \neq P_{A_j}$, there are forbidden combinations and (or) plural representation of number. The last improves diffusion characteristics of the multichannel algorithm on the basis of GPN and serves as an additional barrier on protection of the information against unauthorized access.

For example we represent Fibonacci multichannel algorithm (FMA) and discuss some it's properties. In Table 7 elements $l$-digit Fibonacci PN ($1 \leq l \leq 4$) and it's decimal equivalence are shown.

| N | Decimal number | 4-digit Fibonacci PN | Decimal number | 3-digit Fibonacci PN | Decimal number | 2-digit and 1-digit Fibonacci PN |
|---|---|---|---|---|---|---|
| 1. | 0 | 0000 | 0 | 000 | 0 | 00 |
| 2. | 1 | 0001 | 1 | 001 | 1 | 01 |
| 3. | 1 | 0010 | 1 | 010 | 1 | 10 |
| 4. | 2 | 0011 | 2 | 011 | 2 | 11 |
| 5. | 2 | 0100 | 2 | 100 | 0 | 0 |
| 6. | 3 | 0101 | 3 | 101 | 1 | 1 |
| 7. | 3 | 0110 | 3 | 110 | | |
| 8. | 4 | 0111 | 4 | 111 | | |
| 9. | 3 | 1000 | | | | |
| 10. | 4 | 1001 | | | | |
| 11. | 4 | 1010 | | | | |
| 12. | 5 | 1011 | | | | |
| 13. | 5 | 1100 | | | | |
| 14. | 6 | 1101 | | | | |
| 15. | 6 | 1110 | | | | |
| 16. | 7 | 1111 | | | | |

Table 7. Realization of Fibonacci multichannel algorithm.

From this table it is visible, that growth of Fibonacci weights is small. Therefore, such expressive properties for FMA are existing as plural representation of number and big length for record of number. So, Fibonacci bit recording of $N$-digit alphabet ($A_N$) is bit transformation $\{0,1\}^N \Rightarrow \{0,1\}^M$, where $M \geq N$. Note, that MV2 algorithm and it's clones, binomial multichannel algorithm are based on bit transformation $\{0,1\}^N \Rightarrow \{0,1\}^M$, where $M \leq N$.